\documentclass[aps,prl,reprint,groupedaddress,showpacs,amsmath,amssymb,twocolumn,floatfix]{revtex4-1}
\usepackage{graphicx}
\usepackage{bm}
\usepackage{color}
\usepackage{cleveref}
\usepackage{amsmath}
\usepackage{braket}

\newcommand{\nutotal}{\ensuremath{\nu_{\mathrm{total}}}}

\newcommand{\cg}{\ensuremath{C_{\mathrm{G}}}}
\newcommand{\cp}{\ensuremath{C_{\mathrm{P}}}}
\newcommand{\nut}{\ensuremath{\nu_{\mathrm{t}}}}
\newcommand{\num}{\ensuremath{\nu_{\mathrm{m}}}}
\newcommand{\nub}{\ensuremath{\nu_{\mathrm{b}}}}
\newcommand{\xtm}{\ensuremath{X_{\mathrm{tm}}}}
\newcommand{\xbm}{\ensuremath{X_{\mathrm{bm}}}}
\newcommand{\xtb}{\ensuremath{X_{\mathrm{tb}}}}
\newcommand{\phitm}{\ensuremath{\Phi_{\mathrm{tm}}}}
\newcommand{\phibm}{\ensuremath{\Phi_{\mathrm{bm}}}}
\newcommand{\dlb}{\ensuremath{d/\ell_B}}
\newcommand{\vtg}{\ensuremath{V_{\mathrm{TG}}}}
\newcommand{\vbg}{\ensuremath{V_{\mathrm{BG}}}}
\newcommand{\vbl}{\ensuremath{V_{\mathrm{BL}}}}
\newcommand{\vfast}{\ensuremath{v_{\mathrm{fast}}}}
\newcommand{\vslow}{\ensuremath{v_{\mathrm{slow}}}}
\newcommand{\rhop}{\ensuremath{\rho_{s,+}}}
\newcommand{\rhom}{\ensuremath{\rho_{s,-}}}
\newcommand{\thetap}{\ensuremath{\theta_{s,+}}}

\begin{document}

\title{Electrically Tunable Two-Component Exciton Condensate in a Coulomb-Coupled Graphene Trilayer}

\author{Bo Zou$^{1}$}
\author{A. Okounkova$^{2}$}
\author{Shuaiqing Zhang$^{3}$}
\author{Jian Liao$^{3}$}
\author{J. Pack$^{2}$}
\author{K. Watanabe$^{4}$}
\author{T. Taniguchi$^{4}$}
\author{Yihang Zeng$^{3}$$^{\dag}$}

\affiliation{$^{1}$Department of Physics, University of Michigan, Ann Arbor, Michigan 48109, USA}
\affiliation{$^{2}$Department of Physics, Columbia University, New York, NY 10027, USA}
\affiliation{$^{3}$Department of Physics and Astronomy, Purdue University, West Lafayette, IN 47907, USA}
\affiliation{$^{4}$National Institute for Materials Science, 1-1 Namiki, Tsukuba 305-0044, Japan}

\date{\today}

\maketitle

\textbf{Multicomponent condensates possess internal phase degrees of freedom unavailable to a single-component condensate, yet their components are rarely controllable in solids. Here we realize a graphene trilayer with negligible interlayer tunnelling in which the layer-specific carrier densities are continuously tuned by electrostatic gating. Quantum-capacitance measurements demonstrate that charge-incompressible quantum Hall states at total filling factors $\nu_{\mathrm{total}}=1$ and $2$ persist across the full range of layer-filling configurations and continuously connect the three bilayer exciton-condensate limits. This persistence provides evidence for a trilayer excitonic state. Static Hartree–Fock and time-dependent Hartree–Fock calculations yield two independent finite phase-stiffness eigenmodes and two linearly dispersing Goldstone modes, respectively, when all three layers are partially filled, whereas only one phase-stiffness eigenmodes and one linear Goldstone mode remain when one layer is unfilled. The stiffness eigenmodes rotate continuously between the two adjacent-layer exciton bases as charge is transferred among the layers, revealing electrical control of the condensate-mode composition. Together, the experimental and theoretical results support the identification of a two-component exciton condensate with a continuously tunable internal structure.}

Quantum condensation establishes macroscopic phase coherence and collective transport from a many-body wavefunction described by a complex order parameter. When several condensate components coexist, additional relative phases produce collective modes and topological defects unavailable to a single-component system\cite{moskalenko_1959_2bandsc,Suhl_1959_2bandsc,Leggett_1966_Leggettmode,Mueller_two_component_BEC_2002,Kita_two_component_BEC_1998}. Two-band superconductors provide prominent solid-state examples\cite{zehetmayer_2013_2bandscreview}, but their component densities, couplings and symmetry-breaking fields are largely fixed by material chemistry.  A platform in which the relative populations and composition can be varied continuously would give direct access to the defining internal phase sector of multicomponent order.

Ground-state interlayer excitons provide such an opportunity. Electrons and holes in closely spaced but electrically isolated layers bind through the Coulomb interaction. Recently, two-component exciton condensates formed from distinct spin--valley flavors were identified in an electron--hole bilayer\cite{qi_two-component_2026}. In the quantum Hall regime, a particle-hole transformation maps a partially filled electron layer and a partially empty layer onto interlayer excitons (Fig.~1a), whose condensation spontaneously breaks the relative $U(1)$ symmetry and establishes interlayer phase coherence\cite{Loz.75,Loz.76,shevchenko_theory_1976}. Exciton condensation has been established in GaAs quantum wells\cite{Eis.14,Eis.04,Tutuc.04,Kel.03,Kel.02,Kel.04,Kel.05} and double-layer graphene\cite{Li.17a,Liu.17a,Liu2019,zeng_2026_excitonsolid}. Independent control of the constituent layer densities tunes the emergent exciton density and can drive transitions between excitonic phases, including a superfluid-to-insulator transition\cite{zeng_2026_excitonsolid,zeng_excitondesnitywave_2023,gu_2022_dipolar,zhang_correlated_2022}.

Adding a third Coulomb-coupled layer introduces three pairwise interlayer-coherence channels. When all three layers are partially filled at integer \nutotal, the coherent ground state can break two independent relative U(1) symmetries\cite{hanna1996,ye2005,wang_2025_anyonsuperfluid3layer}. Only two of the three pairwise coherences are independent; we choose the top–middle and middle–bottom channels, \xtm{} and \xbm{}, as a basis for the two-component order parameter (Fig. 1b). These components have independent symmetry-breaking phases and filling-dependent amplitudes. Earlier studies of triple quantum wells found incompressible quantum Hall states when all three layers were populated\cite{jo1992,lay1995}, but lacked independent control of the layer fillings and access to layer-selective response. A layer-addressable trilayer therefore offers a route to identify a two-component condensate and tune its internal composition\cite{hanna1996, ye2005, wang_2025_anyonsuperfluid3layer}.

\begin{figure*}[!t]
\includegraphics[width=0.9\linewidth]{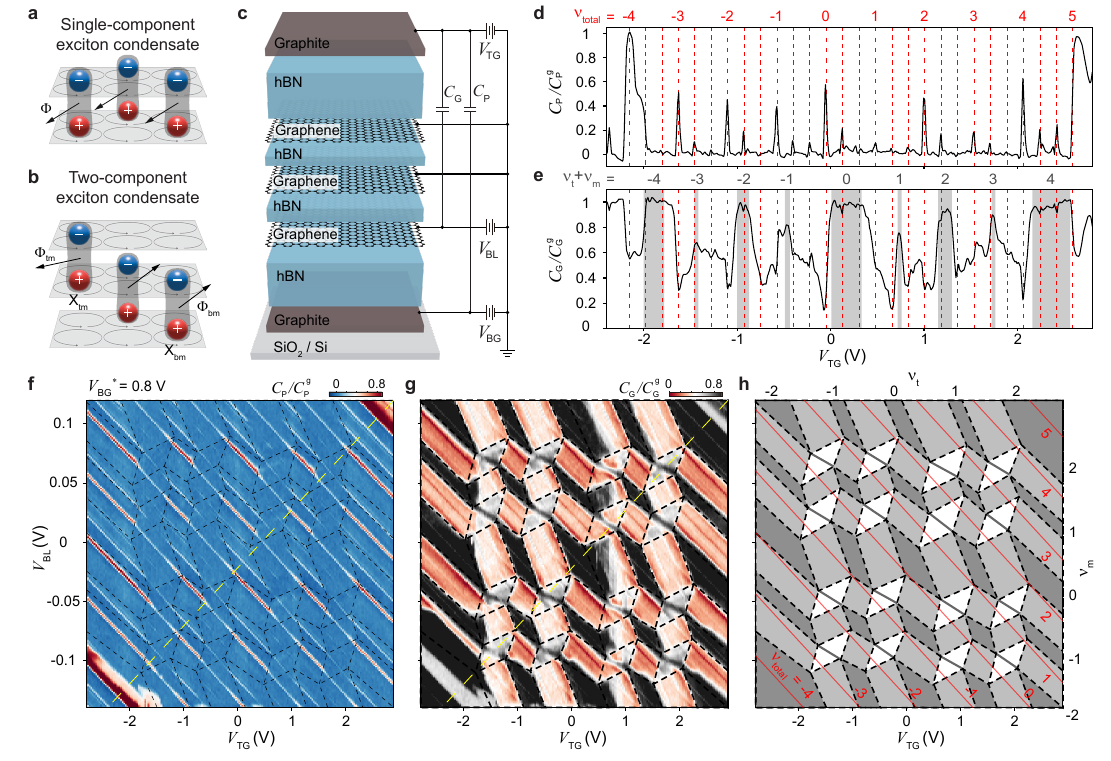}
\caption{\textbf{Coulomb-coupled trilayer graphene.} \textbf{a,b}, Schematics of a single-component exciton condensate in a Coulomb-coupled bilayer (\textbf{a}) and a two-component condensate formed by the adjacent-layer exciton species \xtm{} and \xbm{} with order parameters \phitm{} and \phibm{} in a trilayer (\textbf{b}). \textbf{c}, Device and capacitance-measurement geometry. Three monolayer graphene sheets are separated by 2--3-nm hBN tunnel barriers. The penetration capacitance \cp{} and gate capacitance \cg{} are measured simultaneously at different frequencies. The layer densities are controlled by \vtg, \vbg{} and \vbl. \textbf{d,e}, Normalized \cp{} (\textbf{d}) and \cg{} (\textbf{e}) along the yellow dashed trajectories in \textbf{f} and \textbf{g}, respectively. Peaks in \cp{} occur at integer \nutotal{} (red dashed lines). Gray regions in \textbf{e} mark incompressible states of the top-middle subsystem. \textbf{f,g}, Normalized \cp{} (\textbf{f}) and \cg{} (\textbf{g}) as functions of \vtg{} and \vbl{} at fixed $V_{\mathrm{BG}}^{*}=0.8$~V. Nearly vertical and horizontal dashed boundaries identify layer-resolved integer quantum Hall regions in the top and middle layers, respectively. \textbf{h}, Filling assignment corresponding to \textbf{f,g}. Red diagonal features mark charge-incompressible trilayer states at integer \nutotal{}, as extracted from \textbf{f}. The intersections of the four partially filled symmetry-broken Landau levels in the top and middle layers form a $4\times4$ array of diamond-shaped regions.}
\label{fig:fig1}
\end{figure*}

Here we experimentally realize this setting using three independently contacted graphene monolayers in $N=0$ Landau levels. We observe an incompressible quantum Hall state for arbitrary charge partition among the layers at integer \nutotal{}. When one layer is unfilled, the states reproduce the established bilayer exciton-condensate limits; when all three layers are partially filled, the charge gap remains finite. Mean-field calculations show that the three-layer-filled state supports two linear Goldstone modes and two phase-stiffness channels, whose compositions vary continuously with layer filling. These calculated collective responses provide a theoretical basis for interpreting the measured charge-gapped manifold as a gate-tunable two-component exciton condensate.
At fractional total filling, we also observe incompressible states consistent with coexistence of a fractional quantum Hall state in one layer and a bilayer excitonic state in the other two.

\section{Quantum-capacitance measurements with independent layer control}

The device contains three monolayer graphene sheets separated by 2--3-nm hBN spacers (Fig.~1c), making single-particle tunneling negligible\cite{nguyen_perfect_2025,qi_exciton_2026}. Independent contacts to all three graphene layers allow the top and middle layers to be held at ground potential while the voltages applied to top gate (\vtg), bottom gate (\vbg) and bottom layer (\vbl) provide three independent controls of the layer densities. Ignoring quantum-capacitance corrections, the zeroth-order filling estimates are
\begin{align*}
\nu_{\mathrm{t}}^{0} &= V_{\mathrm{TG}}C_{\mathrm{TG}}/(en_{\phi}),\\
\nu_{\mathrm{m}}^{0} &= V_{\mathrm{BL}}C_{\mathrm{BL}}/(en_{\phi}),\\
\nu_{\mathrm{b}}^{0} &= V_{\mathrm{BG}}^{*}C_{\mathrm{BG}}/(en_{\phi}),
\end{align*}
Unless specified, all capacitances in this paper are expressed per unit area. $C_{\mathrm{TG}}$ and $C_{\mathrm{BG}}$ are the gate-to-graphene geometric capacitances, $C_{\mathrm{BL}}$ is the middle-bottom interlayer geometric capacitance, $V_{\mathrm{BG}}^{*}=V_{\mathrm{BG}}-(1+C_{\mathrm{BL}}/C_{\mathrm{BG}})V_{\mathrm{BL}}$, and $n_{\phi}=eB/h$. Quantum capacitance appreciably redistributes charge among the closely spaced layers, so the layer fillings are calibrated experimentally as described below rather than inferred from these geometric estimates alone. The total filling factor, by contrast, is controlled predominantly by \vtg{} and \vbg{} because $C_{\mathrm{BL}} \gg C_{\mathrm{TG}}, C_{\mathrm{BG}}$, owing to the much thicker hBN gate dielectrics (48 nm and 63 nm) than the spacer hBN between graphene layers.

We measure the gate-to-gate penetration capacitance \cp{} and the top-gate-to-bottom-layer capacitance \cg{} simultaneously (Fig.~1c). To leading order,
\begin{equation*}
C_{\mathrm{P}}\simeq
\frac{C_{\mathrm{P}}^{g}}
{1+C_{Q}^{\mathrm{total}}/(4C_{\mathrm{P}}^{g})},
\end{equation*}
where $C_{\mathrm{P}}^{g}$ is the gate-to-gate geometric capacitance, so \cp{} probes the total quantum capacitance of the trilayer system $C_{Q}^{\mathrm{total}}=e^{2}dn_{\mathrm{total}}/d\mu_{\mathrm{total}}$\cite{ma_exciton_2021_nature}. By contrast,
\begin{equation*}
C_{\mathrm{G}}\simeq
\frac{C_{\mathrm{TG}}C_{Q}^{\mathrm{b}}}
{C_{\mathrm{TG}}+C_{\mathrm{BG}}+C_{Q}^{\mathrm{tm}}+C_{Q}^{\mathrm{b}}},
\end{equation*}, where $C_{Q}^{\mathrm{tm}}$ and $C_{Q}^{\mathrm{b}}$ are the quantum capacitances of the top-middle subsystem and the bottom layer, respectively (see Methods). All data were acquired at $B=15$~T and $T=300$~mK. The limiting cases clarify the layer selectivity of $C_{\mathrm{G}}$. When the bottom layer is incompressible, it cannot accept induced charge and \cg{} is suppressed. When the bottom layer remains compressible but the top-middle subsystem becomes incompressible, reduced screening by the upper subsystem enhances \cg; when both subsystems are compressible, \cg{} takes an intermediate value. These responses therefore provide a direct check of which layer or layer pair is incompressible, independently of the total-compressibility signal in \cp.

Along the trajectory in Fig.~1d, corresponding to $\nu_b^0\approx0.5,\nu_t^0\approx\nu_m^0$, \cp{} exhibits a series of nearly equally spaced peaks, with every third peak particularly prominent. This sequence is consistent with integer quantum Hall states and fractional quantum Hall states (FQHSs) at total fillings $N\pm1/3$. The \cp peaks coincide with local minima along the same trajectory in \cg{} (Fig.~1e). By contrast, maxima in \cg{} identify equally spaced incompressible states of the top-middle subsystem at integer \nut+\num. The yellow dashed lines in Fig.~1f,g indicate the trajectories used for these linecuts. 

The two-dimensional maps in Fig.~1f-h reveal the global phase diagram at fixed zeroth-order bottom-layer filling $\nu_{\mathrm b}^{0}\simeq0.5$. Vertical and horizontal zigzag bands in \cg{} mark integer fillings of the top and middle layers, respectively, and reflect the fourfold-degenerate zeroth Landau level (Fig. 1g,h). The intersections of the four broken symmetry Landau levels (BSLLs) in each layer form a $4\times4$ array of diamond-shaped regions in which both layers are partially filled. Diagonal top-middle incompressible states traverse these diamonds, as expected for bilayer exciton condensates that remain charge gapped under continuous charge transfer\cite{Li.17a,Liu.17a,Li2019,liu_interlayer_2019,Liu2019,zeng_2026_excitonsolid}. By contrast, the integer-\nutotal{} features in \cp{} remain nearly straight while crossing regions with strongly varying layer compressibility and charge distribution (Fig. 1f), confirming that the total filling factor is largely insensitive to quantum-capacitance corrections. Their persistence motivates a systematic study of layer-filling space for which we focus on the first BSLL in each layer ($0\leq\nu_t,\nu_m,\nu_b\leq1$).

\section{Incompressibility throughout layer-filling space}

\begin{figure*}[!t]
\centering
\includegraphics[width=0.8\linewidth]{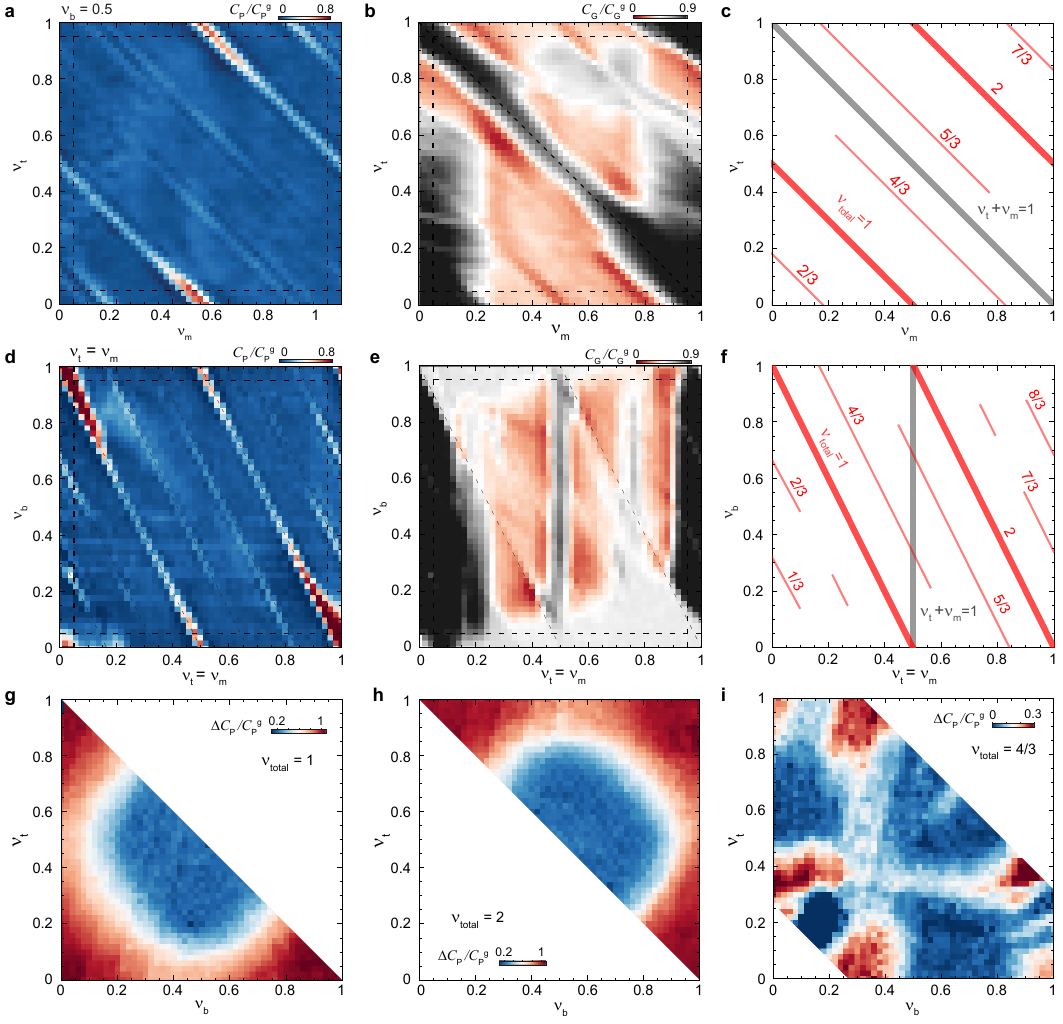}
\caption{\textbf{Incompressible quantum Hall states across layer-filling space.} \textbf{a,b}, Normalized penetration capacitance \(C_{\mathrm P}\) (\textbf{a}) and gate capacitance \(C_{\mathrm G}\) (\textbf{b}) at fixed \(\nu_b=0.5\). \textbf{c}, Guide to the incompressible-state assignments in \textbf{a,b}. \textbf{d,e}, Normalized \(C_{\mathrm P}\) (\textbf{d}) and \(C_{\mathrm G}\) (\textbf{e}) along \(\nu_t=\nu_m\). \textbf{f}, Guide to the assignments in \textbf{d,e}. In \textbf{c,f}, thick red lines denote integer \(\nu_{\mathrm{total}}\), thin red lines denote fractional \(\nu_{\mathrm{total}}\), and gray lines denote the top--middle bilayer state at \(\nu_t+\nu_m=1\). \textbf{g-i}, Normalized penetration-capacitance peak amplitude \(\Delta C_{\mathrm P}/C_{\mathrm P}^{g}\) at \(\nu_{\mathrm{total}}=1\) (\textbf{g}), \(2\) (\textbf{h}), and \(4/3\) (\textbf{i}),  plotted as functions of \(\nu_t\) and \(\nu_b\), with \(\nu_m=\nu_{\mathrm{total}}-\nu_\mathrm{t}-\nu_\mathrm{b}\). White regions in \textbf{g--i} lie outside the physical range \(0\leq\nu_m\leq1\). }
\label{fig:fig2}
\end{figure*}

Following previous studies of quantum Hall bilayers~\cite{Li2019,liu_interlayer_2019,zhang_excitons_2025}, we obtain more accurate layer-specific filling factors (\nut, \num, \nub) from $(\nu_{\mathrm{t}}^0,\nu_{\mathrm{m}}^0,\nu_{\mathrm{b}}^0)$ using a linear transformation calibrated against several incompressibility features observed in \cp{} and \cg{} (Supplementary Information). 
Figure~2 shows two representative scans of the three-dimensional filling dependence. At fixed $\nub=0.5$, the \cp{} map contains persistent incompressible trajectories at \nutotal= 1 and 2, together with intermittent trajectories at \nutotal= $N\pm1/3$. Other representative slices at fixed \nub{} show qualitatively similar incompressible states in \cp (Extended data Fig 2). Along $\nut=\num$, a complementary scan changes the top and middle fillings together and the bottom filling independently, also revealing persistent integer-\nutotal{} incompressible states (Fig.~2d). The corresponding \cg{} maps (Fig.~2b,e) locate the top-middle bilayer incompressible quantum Hall state along $\nut+\num=1$, as expected. Figure~2c,f summarize the experimentally observed incompressible states in \cp{} (red) and \cg{} (gray). 

We next follow the \nutotal$=1$ and 2 incompressible peaks while varying the charge partition among all three layers. The normalized peak amplitude $\Delta C_{\mathrm{P}}/C_{\mathrm{P}}^{g}$ remains greater than $0.1$ throughout both filling triangles (Fig.~2g,h), bounded by $0\leq\nu_t,\nu_m,\nu_b\leq1$. The local background and peak-extraction procedure are defined in Extended Data Fig.~1. In the absence of interlayer correlations, three partially filled Landau levels would generally be compressible. Together with negligible tunnelling, the persistent integer-total-filling incompressibility provides thermodynamic evidence for a trilayer charge-gapped state driven by interlayer Coulomb interaction. Under a particle-hole transformation, these trilayer charge-gapped states correspond to excitonic states involving all three layers. 

In contrast to the persistent integer quantum Hall states, the \nutotal=4/3 incompressible features are concentrated near \nut=1/3 and \nub=1/3, with additional localized maxima at (\nut,\num,\nub)=(0,1/3,1), (1,1/3,0) and (2/3,0,2/3) (Fig.~2i). Along \nut=1/3, the two maximally incompressible endpoints occur at (1/3,0,1) and (1/3,1,0), the ridge interior is therefore consistent with a 1/3-filled top layer coexisting with an \xbm{} condensate in the other two layers. An analogous scenario applies along \nub= 1/3. The intersection of the two incompressible ridges at (1/3,,2/3,,1/3) is particularly intriguing. At the experimentally relevant adjacent-layer \dlb{}, strong Coulomb coupling favors excitonic ground states over layer-decoupled FQHSs \cite{zeng_2026_excitonsolid}. At the intersection, however, the top–middle and middle–bottom pairing channels draw on the same middle-layer degrees of freedom and may therefore compete. This competition could select one pairing channel, produce a coherent state involving both, or suppress conventional interlayer coherence in favor of a distinct three-layer-correlated incompressible phase. Determining which possibility is realized will require further experimental and theoretical study. A similar phase diagram is observed at \nutotal=5/3 (Extended data Fig. 3), as expected from the layer-resolved particle-hole transformation (\nut,\num,\nub)$\leftrightarrow$(1-\nut,1-\num,1-\nub). The fractional-\nutotal{} incompressible features at the expected layer-specific commensurate fillings provides an independent consistency check of the calibrated layer-specific filling-factor assignment.


\section{Charge Gap and Goldstone modes}

\begin{figure*}
\includegraphics[width=0.8\linewidth]{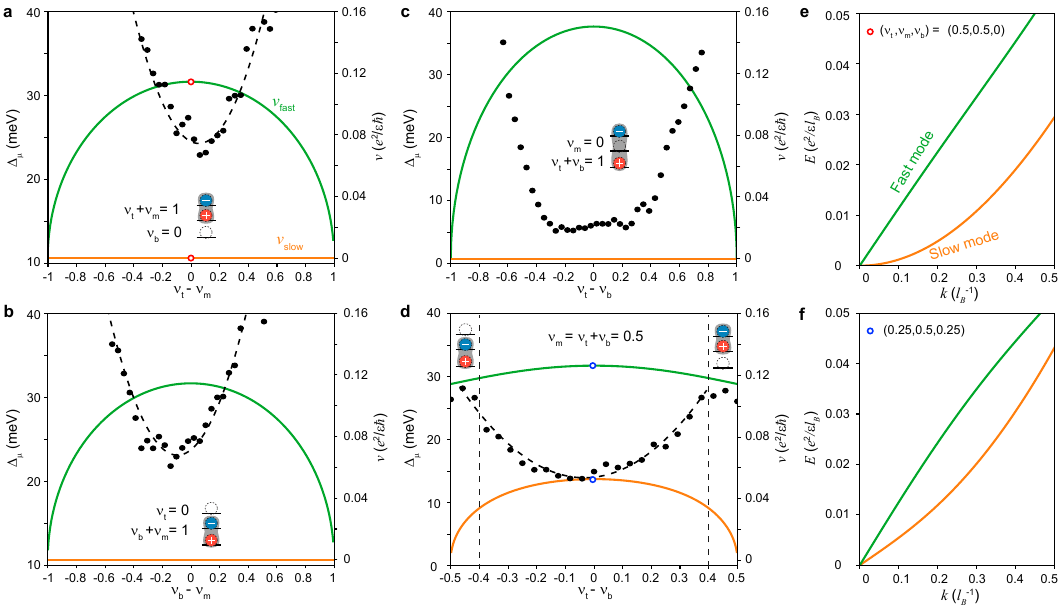}
\caption{\textbf{Charge gap and neutral collective modes at \nutotal = 1.} \textbf{a-d}, Measured thermodynamic charge gap $\Delta_\mu$ (black, left axis) and calculated sound velocities (right axis) along the top-middle bilayer limit \nub=0 (\textbf{a}), middle-bottom bilayer limit \nut\ =0 (\textbf{b}), top-bottom bilayer limit \num=0 (\textbf{c}), and \num=\nut+\nub=0.5 (\textbf{d}). Green and orange curves show the fast- and slow-mode velocity, \vfast and \vslow, respectively. The dashed parabolas are guides to the eye; in \textbf{d}, vertical black dashed lines mark crossovers into the gap-saturated bilayer-condensate regimes. \textbf{e,f}, Calculated neutral collective mode dispersions at $(\nut,\num,\nub)=(0.5,0.5,0)$ (\textbf{e}) and $(0.25,0.5,0.25)$ (\textbf{f}).}
\end{figure*}

The thermodynamic charge gap is extracted from the area of each baseline-subtracted \cp{} peak along a trajectory perpendicular to the constant \nutotal{} plane,
$\Delta_\mu\approx \frac{e^2}{4C_{\mathrm{P}}^{g}}\frac{eB}{h}\int \frac{\Delta C_{\mathrm{P}}}
{C_{\mathrm{P}}^{g}}\,d\nutotal$ 
, as illustrated in Extended Data Fig.~1~\cite{Eisenstein_prl_1992_compressibility,ma_exciton_2021_nature}. When $\nub=0$, the \nutotal$=1$ state is a condensate of top-middle excitons \xtm{}, and its charge gap is minimized near balanced filling (Fig.~3a). When $\nut=0$, the state similarly realizes a condensate of middle-bottom excitons \xbm{} and shows a comparable filling dependence (Fig.~3b). In these adjacent-layer limits, the gap follows an approximately quadratic dependence on filling imbalance, consistent with charge gap measurements in quantum Hall bilayers \cite{Tutuc_imbalance_2003}. When $\num=0$, the more widely separated top and bottom layers form \xtb{} excitons; their condensate has a broader and lower charge gap minimum and deviates more strongly from the quadratic dependence (Fig.~3c). This deviation may arise from stronger interactions between excitons with larger dipole moments. The persistence of a charge gap in all three bilayer limits confirms that the Coulomb interaction is sufficiently strong to sustain an excitonic state between each pair of occupied layers. 

Figure~3d follows a continuous trajectory between the two adjacent-layer exciton-condensate limits along $\num=\nut+\nub=0.5$. Transferring charge from the bottom to the top layer carries the system from the \xbm\ condensate limit at $(0,0.5,0.5)$ to the \xtm condensate limit at $(0.5,0.5,0)$. Outside the vertical black dashed lines, the charge gap saturates at its bilayer-limit value as one outer layer becomes dilute. Between them, the charge gap remains finite and evolves continuously through the three-layer-filled region, again following an approximately quadratic dependence (dashed lines). Together with the negligible direct interlayer hybridization, this continuous evolution supports a trilayer excitonic state continuously connected to the two adjacent-layer condensate limits.

\begin{figure*}
\includegraphics[width=0.8\linewidth]{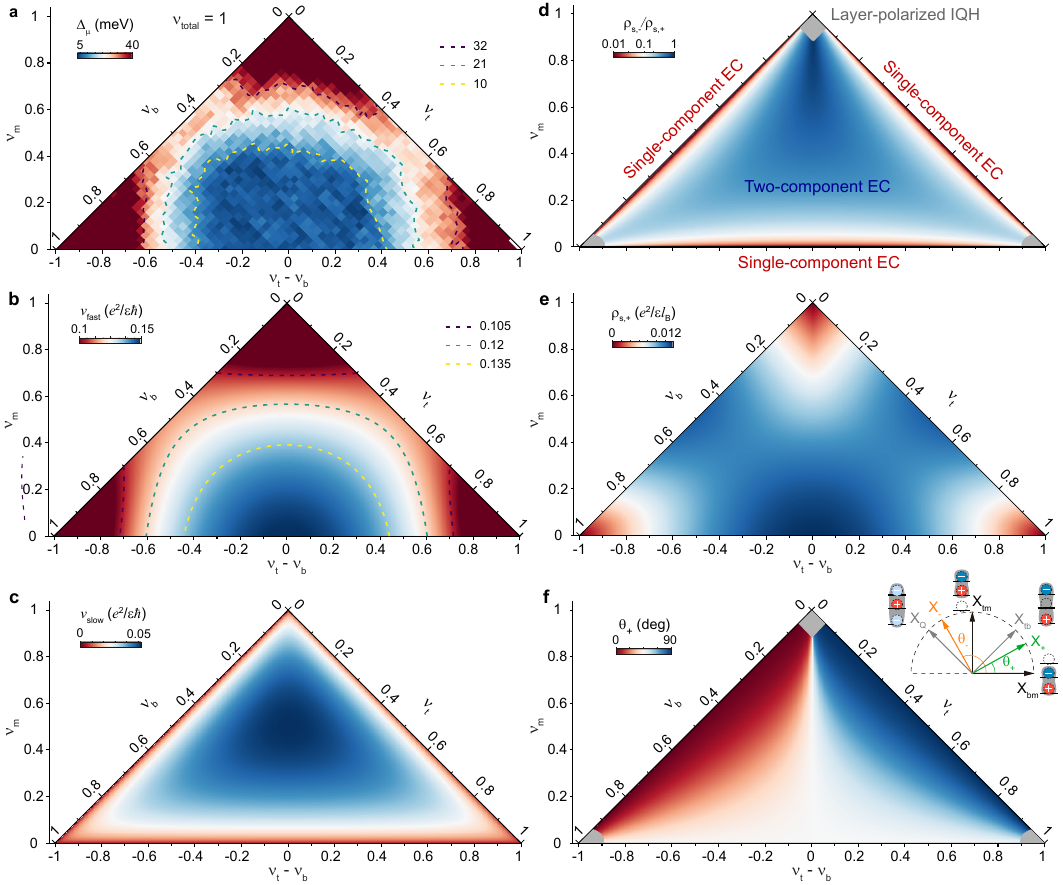}
\caption{\textbf{Filling factor phase diagram and electrical tunability.} \textbf{a}, Experimentally measured thermodynamic charge gap of the \nutotal$=1$ state versus layer fillings. Gap values are displayed only up to the operational quantitative limit of \(40\) meV. Saturated regions denote nominal extractions above this limit. \textbf{b,c}, Calculated velocities of the fast (\textbf{b}) and the slow Goldstone mode (\textbf{c}) from TDHF. Dashed contours in \textbf{a,b} guide comparison of their opposite filling dependence. The color scale in \textbf{b} is clipped in the lower limit. See Extended Data Fig. 5 for plot with an unsaturated color scale. \textbf{d}, Ratio of the secondary to principal phase-stiffness eigenvalues, $\rhom/\rhop$, plotted on a logarithmic color scale. Values below 0.01 are clipped to the lower color limit. \textbf{e,f}, Principal phase stiffness mode eigenvalue \rhop (\textbf{e}) and its eigenvector angle \thetap{} (\textbf{f}) in the $(\xtm,\xbm)$ basis. Layer-polarized corners, which have no Goldstone modes and where \rhop{} falls below 20\% of its maximum value, are masked in gray in \textbf{d} and \textbf{f}.}
\end{figure*}

To determine the ground-state order and neutral collective structure of the trilayer excitonic state at \nutotal$=1$, we model the system as three spin- and valley-polarized zeroth Landau levels coupled by gate-screened Coulomb interactions. We assume equal adjacent-layer separations, motivated by the similar behavior of the two adjacent-layer limits in Fig.~3a,b. Static Hartree--Fock (HF) calculations yield the interlayer-coherent ground state and its phase-stiffness matrix, whereas time-dependent Hartree--Fock (TDHF) calculations yield the neutral collective-mode dispersions (Supplementary Information).

At $(\nut,\num,\nub)=(0.5,0.5,0)$, where the bottom layer is empty, TDHF yields one linearly dispersing Goldstone mode and a second branch with quadratic long-wavelength dispersion and hence vanishing velocity as $k\rightarrow0$ (Fig.~3e). The presence of only one linear Goldstone mode is consistent with a single-component exciton condensate in this effective bilayer limit\cite{Moon_1995_meron,hanna1996}. By contrast, at $(\nut,\num,\nub)=(0.25,0.5,0.25)$, TDHF yields two linearly dispersing Goldstone modes (Fig.~3f), as expected for a two-component condensate. The calculated sound velocities \vfast{} and \vslow{} are overlaid on the measured gap linecuts in Fig.~3a--d. Along both the effective bilayer limits and the three-layer linecut, the thermodynamic gap shows a strong anticorrelation with \vfast{}. This phenomenological anticorrelation suggests that the charged excitations and the fast neutral mode may share a common dependence on exchange and capacitive energetics.

\section{Continuous tuning of the two-component exciton condensate}

The full filling-space maps reveal the relation between experiment and theory (Fig.~4). The measured \nutotal$=1$ charge gap persists throughout the triangle defined by $0\leq \nut,\num,\nub\leq1$. The gap is largest near the layer-polarized integer quantum Hall corners and exhibits a broad minimum near the center of the $\num=0$ edge. The \nutotal$=2$ gap has a similar structure, as expected from the particle-hole symmetry (Extended Data Fig.~4).

The TDHF results distinguish the number of linearly dispersing Goldstone modes across filling
space. The fast-mode velocity remains finite along all three effective bilayer limits and vanishes only at the layer-polarized corners (Fig. 4b; see Extended Data Fig. 5a for an unsaturated color scale). The slow-mode velocity is finite only in the three-layer-filled region and approaches zero along the three edges as any layer is depleted (Fig.~4c). Thus, within TDHF, the three-layer-filled region supports two linear Goldstone modes, as expected for a two-component exciton condensate, whereas each effective bilayer edge supports only one, as expected for a single-component condensate. The opposing filling dependences of the experimental charge gap and \vfast{} extends the phenomenological anticorrelation observed in Fig. 3 linecuts across the full phase diagram, as emphasized by the corresponding contours in Fig.~4a,b.

The static HF calculation provides a complementary characterization. Phase twists of the adjacent-layer condensate components $(\xtm,\xbm)$ are governed by a real symmetric $2\times2$ stiffness matrix with eigenvalues $\rhop\geq\rhom\geq0$. Its eigenvectors define the principal and secondary phase-stiffness modes as linear combinations of the \xtm{} and \xbm{} components, while the eigenvalues quantify their phase rigidity. 
The principal stiffness \rhop{} remains finite away from the layer-polarized corners (Fig.~4e, Extended Data Fig. 5). Within the three-layer filled interior, \rhom{} is also finite, and a nonzero $\rhom/\rhop$ therefore identifies the rank-two phase-stiffness regime (Fig. 4d). Along the effective bilayer edges, \rhom{} vanishes and the stiffness matrix has rank one.

The principal condensate phase mode rotates smoothly from predominantly \xtm{} character near one adjacent-layer limit to predominantly \xbm{} character near the opposite limit, as quantified by the condensate-composition angle \thetap{} (Fig.~4f). Wherever both stiffness eigenvalues are finite, the secondary mode
is orthogonal to the principal mode (Extended Data Fig. 5). Along the layer-symmetric trajectory $\nut=\nub$, symmetry gives equal \xtm{} and \xbm{} weights. Their in-phase combination produces a net top-to-bottom dipolar phase channel equivalent to that of a \xtb{} condensate, whereas their out-of-phase combination produces an internal quadrupolar channel with no net top-to-bottom dipole (Fig.~4f inset). Moving away from these trajectories continuously transfers condensate-mode weight between \xtm{} and \xbm{}. 

Together, the persistent charge gap, the two calculated Goldstone modes and the rank-two phase-stiffness structure support the identification of a two-component exciton condensate at \nutotal=1 and, by particle–hole symmetry, at \nutotal=2, when all three layers are partially filled. Both the calculated condensate-mode composition and measured charge gap vary continuously with layer filling and connect smoothly to the single-component exciton-condensate limit. These results reveal an electrically tunable two-component condensate.

Direct layer-resolved Coulomb-drag measurements could next probe the two independent neutral channels. A layer-configuration-dependent current-drag ratio or Hall response would probe the gate-tuned collective-mode structure, while a two-step transition in counterflow resistance could resolve the two condensate modes. More broadly, Coulomb-coupled multilayers provide a controlled setting for multicomponent phase coherence, composite topological excitations and new anyonic quasiparticles~\cite{wang_2025_anyonsuperfluid3layer}. With the layer degree of freedom interpreted as a tunable pseudospin, such multilayer systems provide a route to simulating high-spin quantum systems with electrically controlled populations and interactions.

\bibliography{YZ_ref}

\section{Methods}

\subsection{Device fabrication}
The Coulomb-coupled graphene trilayer heterostructure was assembled using the van der Waals dry-transfer technique as described in ref.~\cite{Zeng2018}. The region containing both graphite gates and all three graphene layers was defined by electron-beam lithography and reactive-ion etching. Cr/Pd/Au was deposited on the exposed, non-overlapping edges of the graphene and graphite layers to form layer-specific one-dimensional contacts\cite{Dean.10}. The thicknesses of the two hBN tunnel barriers were measured by atomic force microscopy before assembly.

\subsection{Capacitance measurements}
We followed the quantum-capacitance technique of ref.~\cite{shi_bilayer_2022}. Excitations at 37.77~kHz and 17.77~kHz were applied to the bottom gate and bottom graphene layer to measure \cp{} and \cg, respectively. The induced top-gate signals were amplified by a low-temperature high-electron-mobility transistor and demodulated at room temperature using separate lock-in amplifiers. The complex response was rotated so that the dissipative channel was approximately zero in compressible regions. For \cp, a constant parasitic background was subtracted such that the capacitive response approached zero at high density ($|\nu|>2$). The geometric penetration capacitance $C_{\mathrm{P}}^{g}$ was obtained from the saturation value when all three layers were in layer-polarized integer quantum Hall states. The \cg{} channel was processed in the same way except that no constant background was subtracted because it lacks an unambiguous zero-response region (Extended Data Fig.~6).

\subsection{Interpretation of the gate capacitance}
In the small-signal circuit, the bottom graphene layer is driven, the induced charge is measured on the top gate, and the top graphene layer, middle graphene layer and bottom gate are held at fixed electrochemical potential. Denoting the electrostatic potentials of the top-middle subsystem and bottom layer by $\phi_{\mathrm{tm}}$ and $\phi_{\mathrm{b}}$, respectively, the linear response gives
\begin{align*}
(C_{\mathrm{TG}}+C_{\mathrm{BL}}+C_{Q}^{\mathrm{tm}})\phi_{\mathrm{tm}}
-C_{\mathrm{BL}}\phi_{\mathrm{b}} &= C_{\mathrm{TG}}V_{\mathrm{TG}},\\
-C_{\mathrm{BL}}\phi_{\mathrm{tm}}
+(C_{\mathrm{BL}}+C_{\mathrm{BG}}+C_{Q}^{\mathrm{b}})\phi_{\mathrm{b}} &=0.
\end{align*}
By mutual-capacitance reciprocity, this geometry is equivalent to driving the top gate and measuring the induced charge on the bottom graphene layer; we use this reciprocal representation below. The measured mutual capacitance per unit area is
\begin{align*}
C_{\mathrm{G}}
&=-\frac{\partial q_{\mathrm{b}}}{\partial V_{\mathrm{TG}}}
=\frac{C_{\mathrm{TG}}C_{\mathrm{BL}}C_{Q}^{\mathrm{b}}}{\mathcal D},\\
\mathcal D
&=(C_{\mathrm{TG}}+C_{\mathrm{BL}}+C_{Q}^{\mathrm{tm}})
(C_{\mathrm{BL}}+C_{\mathrm{BG}}+C_{Q}^{\mathrm{b}})
-C_{\mathrm{BL}}^{2}.
\end{align*}
Because $C_{\mathrm{BL}}\gg C_{\mathrm{TG}},C_{\mathrm{BG}}$ in this device,
\begin{equation*}
C_{\mathrm{G}}\simeq
\frac{C_{\mathrm{TG}}C_{Q}^{\mathrm{b}}}
{C_{\mathrm{TG}}+C_{\mathrm{BG}}+C_{Q}^{\mathrm{tm}}+C_{Q}^{\mathrm{b}}}.
\end{equation*}
Thus \cg{} is suppressed when the bottom layer is incompressible and enhanced when the bottom layer is compressible while the top-middle subsystem is incompressible.

\section{Data availability}
The data supporting the plots and other findings of this study and the data analysis codes are available in Purdue University Research Repository (PURR).

\section{Acknowledgements}
Y.Z. and B.Z. acknowledge helpful discussions with Qi Zhou, Yahui Zhang, Yanqi Wang, Allan H. Macdonald, Qianhui Shi, Jia Li and Cory Dean. K.W. and T.T. acknowledge support from the MEXT Element Strategy Initiative to Form Core Research Center (grant JPMXP0112101001) and CREST, JST (JPMJCR15F3).

\section{Author contributions}
Y.Z. conceived the experiment. A.O. and Y.Z. fabricated the samples. Y.Z. and J.P. performed the capacitance measurements. Y.Z., S.Z. and J.L. performed data analysis. B.Z. performed the Hartree-Fock calculations. B.Z. and Y.Z. wrote the manuscript with input from all authors.

\section*{Competing interests}
The authors declare no competing interests.

\renewcommand{\figurename}{Extended Data Fig.}
\setcounter{figure}{0}

\begin{figure*}
\includegraphics[width=1\linewidth]{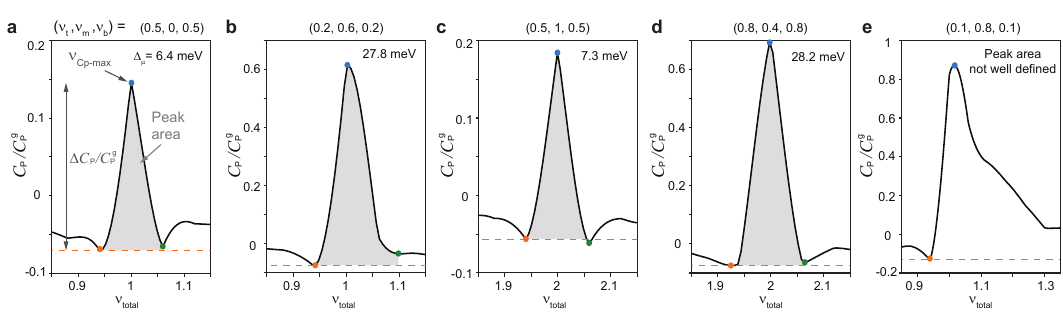}
\caption{\textbf{Extraction of peak position, peak amplitude and thermodynamic charge gap.} Representative $C_{\mathrm{P}}/C_{\mathrm{P}}^{g}$ linecuts taken perpendicular to the \nutotal$=1$ or 2 plane in $(\nut,\num,\nub)$ space. The peak position $\nu_{\mathrm{Cp,max}}$ is identified first (blue circles). The nearest minima within $\pm0.12$ in \nutotal{} on the low- and high-filling sides are then identified (orange and green circles). The better-defined low-filling minimum sets the local baseline (orange dashed line). The negative baseline reflects the negative-compressibility background of the lowest Landau level. Using the mean of the two minima or a constant baseline does not change the qualitative filling dependence. The excess peak amplitude is the difference between the maximum and the baseline. The thermodynamic gap is obtained by integrating the baseline-subtracted peak between the two minima (shaded regions) and applying the filling-to-gate-voltage conversion $\Delta_\mu\approx \frac{e^2}{4C_{\mathrm{P}}^{g}}\frac{eB}{h}\int \frac{\Delta C_{\mathrm{P}}}
{C_{\mathrm{P}}^{g}}\,d\nutotal=~(0.61~\mathrm{eV})\int \frac{\Delta C_{\mathrm{P}}}
{C_{\mathrm{P}}^{g}}\,d\nutotal$. \textbf{a,c} Linecuts at the centers of the least incompressible regions for $\nutotal=1$ and 2, respectively. \textbf{b,d} Representative linecuts with larger charge gaps of around 30 meV for $\nutotal=1$ and 2, respectively. \textbf{e}, Representative linecut from the large-incompressibility regime near a layer-polarized integer quantum hall state. Although the incompressible peak remains evident, the neighboring compressible response does not fully equilibrate at the measurement frequency. Consequently, the integration range and thermodynamic charge gap cannot be determined quantitatively. Based on the largest gaps across the full dataset for which both neighboring compressible minima remain clearly resolved, we adopt 40~meV as the operational upper limit for quantitative charge-gap extraction.}
\end{figure*}


\begin{figure*}
\includegraphics[width=0.95\linewidth]{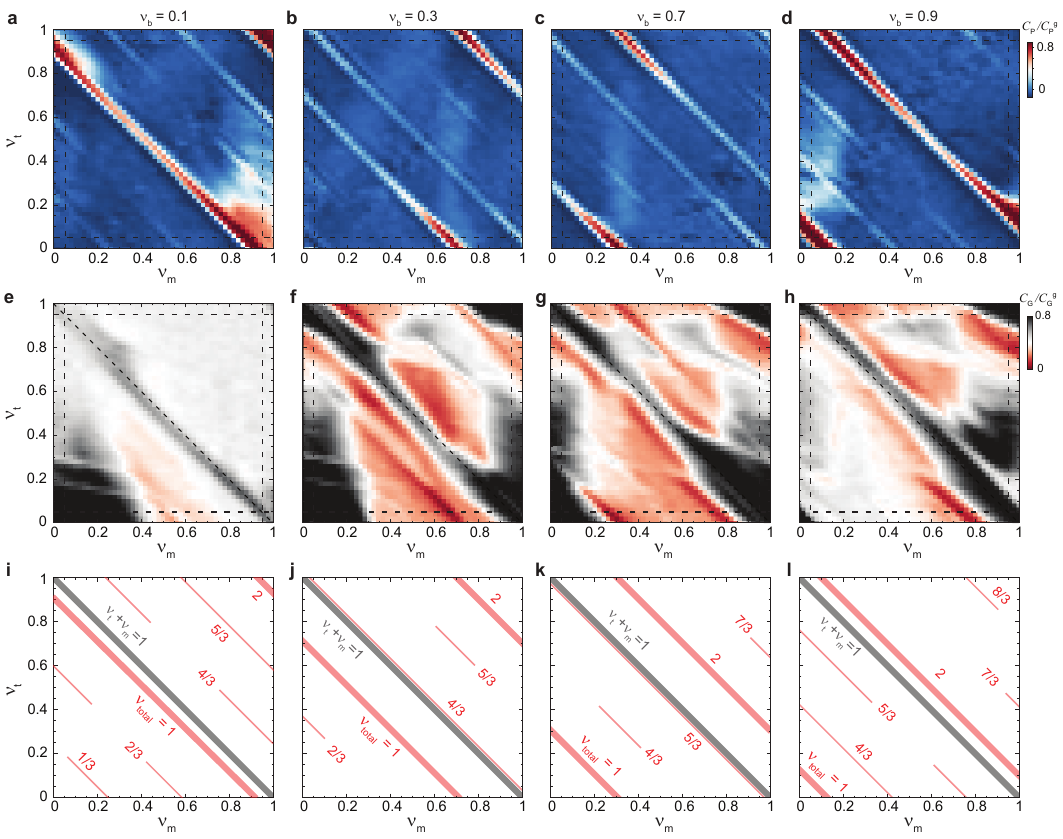}
\caption{\textbf{Layer-resolved filling-factor calibration from incompressible states.} \textbf{a--d}, Normalized penetration-capacitance maps, $C_{\mathrm{P}}/C_{\mathrm{P}}^{g}$, as functions of the top- and middle-layer filling factors, $\nu_{\mathrm{t}}$ and $\nu_{\mathrm{m}}$, at fixed bottom-layer fillings $\nu_{\mathrm{b}}=0.1$, 0.3, 0.7 and 0.9, respectively. The diagonal incompressible features occur at integer and fractional values of the total filling $\nu_{\mathrm{total}}=\nu_{\mathrm{t}}+\nu_{\mathrm{m}}+\nu_{\mathrm{b}}$ and therefore shift with $\nu_{\mathrm{b}}$.
\textbf{e--h}, Corresponding normalized gate-capacitance maps, $C_{\mathrm{G}}/C_{\mathrm{G}}^{g}$, at the same fixed values of $\nu_{\mathrm{b}}$. The prominent diagonal feature follows $\nu_{\mathrm{t}}+\nu_{\mathrm{m}}=1$ and remains fixed as $\nu_{\mathrm{b}}$ is varied, identifying the top--middle exciton condensate. The dashed diagonal lines indicate $\nu_{\mathrm{t}}+\nu_{\mathrm{m}}=1$.
\textbf{i--l}, Schematic summaries of the features observed in the corresponding $C_{\mathrm{P}}$ and $C_{\mathrm{G}}$ maps. Gray lines denote the top--middle condensate at $\nu_{\mathrm{t}}+\nu_{\mathrm{m}}=1$, whereas red lines denote incompressible states at the indicated values of $\nu_{\mathrm{total}}$.}
\end{figure*}

\begin{figure*}
\includegraphics[width=0.85\linewidth]{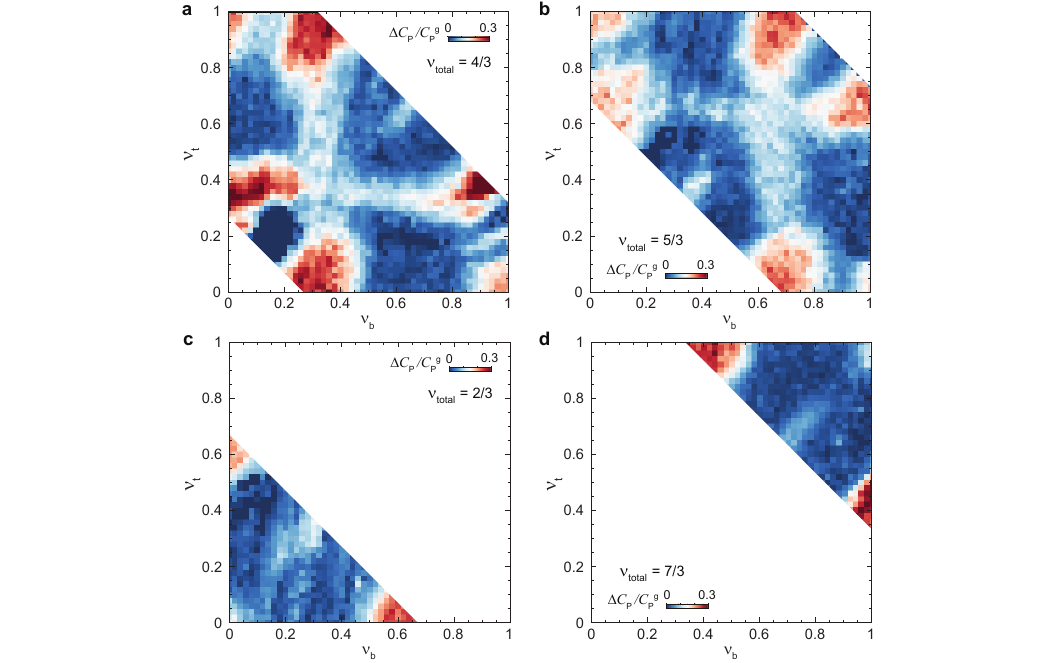}
\caption{\textbf{Layer-filling dependence of incompressibility at fractional \nutotal.} \textbf{a,b}, Normalized \cp{} peak at $\nutotal=4/3$ and 5/3.The isolated incompressible regions correspond to layer-decoupled quantum Hall states in which each layer has a specific fractional or integer filling. The vertical and horizontal incompressible stripes are consistent with coexistence of a bilayer exciton condensate in two layers and a 1/3 or 2/3 FQHS in the third. \textbf{c,d},  Normalized \cp{} peak amplitude at \nutotal=2/3 and 7/3 showing incompressibility only when each layer is at a specific filling. The pairs $\nutotal=4/3, 5/3$ and $\nutotal=2/3,7/3$ show similar patterns under particle-hole symmetry.}
\end{figure*}

\begin{figure*}
\includegraphics[width=0.85\linewidth]{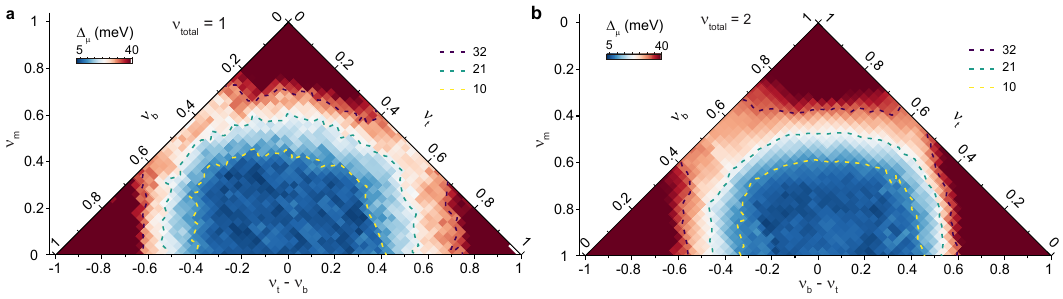}
\caption{\textbf{Layer-filling dependence of the thermodynamic gap at \nutotal$=1$ and 2.} \textbf{a,b}, Thermodynamic charge gap at \nutotal$=1$ (\textbf{a}, reproduced from Fig.~4a) and \nutotal$=2$ (\textbf{b}). The two maps show similar filling dependences under the particle-hole transformation $(\nut,\num,\nub)\leftrightarrow(1-\nut,1-\num,1-\nub)$.}
\end{figure*}

\begin{figure*}
\includegraphics[width=0.85\linewidth]{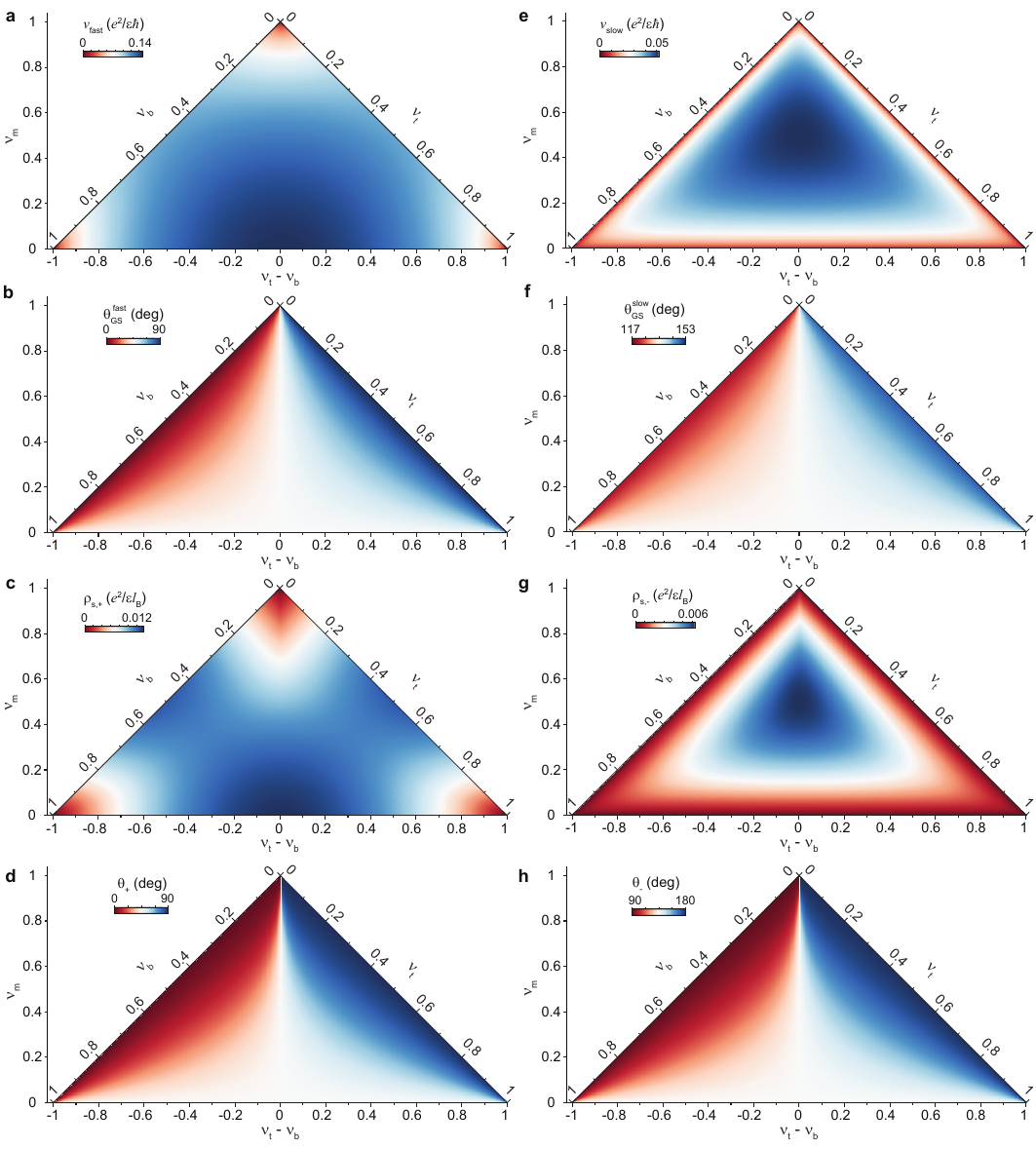}
\caption{\textbf{Calculated dynamical and static properties of the interlayer-coherent state.} \textbf{a,e}, TDHF velocities of the fast (\textbf{a}) and slow (\textbf{e}) Goldstone branches. \textbf{b,f}, TDHF eigenvector angles of the fast (\textbf{b}) and slow (\textbf{f}) branches, defined from the layer-density response as described in the Supplementary Information. \textbf{c,g}, Principal (\textbf{c}) and secondary (\textbf{g}) eigenvalues of the static HF phase-stiffness matrix. \textbf{d,h}, Corresponding HF stiffness-eigenvector angles in the $(\xtm,\xbm)$ basis. The TDHF and HF angles characterize different eigenproblems and are not generally identical.}
\end{figure*}

\begin{figure*}
\includegraphics[width=0.9\linewidth]{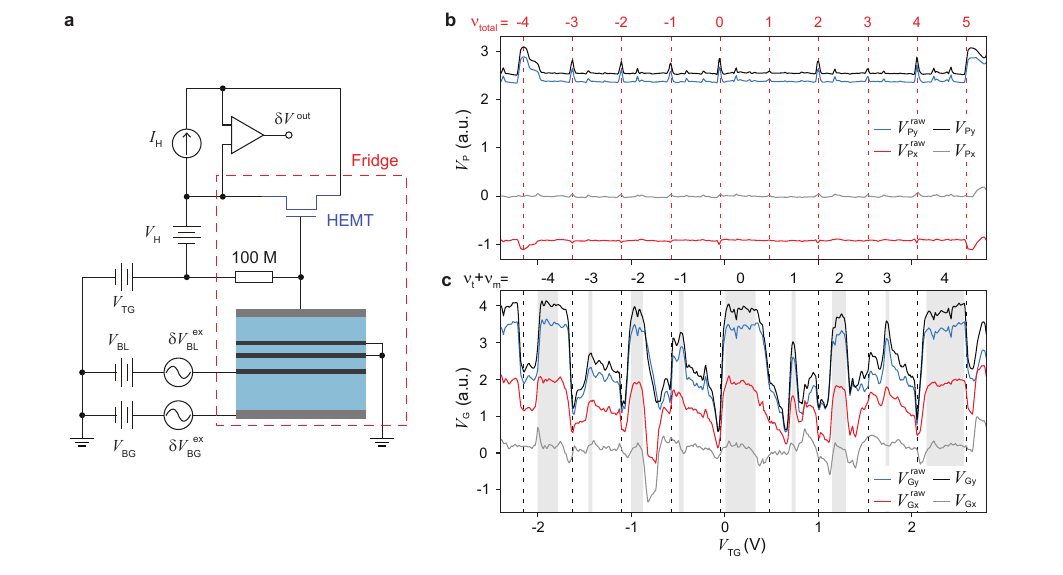}
\caption{\textbf{Capacitance measurement and signal processing.} \textbf{a}, Measurement circuit, following ref.~\cite{shi_WSe2_FQHE_2020}. \textbf{b,c}, Raw and rotated complex responses measured along the same trajectory as Fig.~1d,e. The two lock-in channels are rotated so that the dissipative component is approximately zero in compressible regions, yielding capacitive (black) and dissipative (gray) responses. The parasitic background of \cp{} is subtracted such that \cp{} approaches zero when all layers occupy high Landau levels. $C_{\mathrm{P}}^{g}$ and $C_{\mathrm{G}}^{g}$ are the corresponding saturation values used for normalization.}
\end{figure*}

\end{document}



\begin{center}
{\large\bfseries Supplementary Information for\\[0.4em]
``Electrically Tunable Two-Component Exciton Condensate in a Coulomb-Coupled Graphene Trilayer''\par}
\end{center}

\author{Bo Zou$^{1}$}
\author{A. Okounkova$^{2}$}
\author{Shuaiqing Zhang$^{3}$}
\author{Jian Liao$^{3}$}
\author{J. Pack$^{2}$}
\author{K. Watanabe$^{4}$}
\author{T. Taniguchi$^{4}$}
\author{Yihang Zeng$^{3}$$^{\dag}$}

\affiliation{$^{1}$Department of Physics, University of Michigan, Ann Arbor, Michigan 48109, USA}
\affiliation{$^{2}$Department of Physics, Columbia University, New York, NY 10027, USA}
\affiliation{$^{3}$Department of Physics and Astronomy, Purdue University, West Lafayette, IN 47907, USA}
\affiliation{$^{4}$National Institute for Materials Science, 1-1 Namiki, Tsukuba 305-0044, Japan}

\date{\today}
\maketitle

\noindent\textbf{This PDF file includes:}

\noindent{Supplementary Text}

\noindent{Materials and Methods}



\renewcommand{\thefigure}{S\arabic{figure}}
\setcounter{figure}{0}
\setcounter{equation}{0}

\vspace{10pt}

\section{Determination of filling factors}
Our device consists of three monolayer graphene sheets separated by two 2--3-nm-thick hBN crystals. The much thicker top- and bottom-gate hBN dielectrics have thicknesses $d_\text{TG}= 48$ nm and $d_\text{BG}= 63$ nm, respectively, as determined from the Landau-level spacing. Each graphene layer is contacted independently, allowing an interlayer bias to be applied. The top and middle graphene layers are held at ground potential throughout the experiment. Considering only the geometric capacitances, the zeroth-order estimates of the layer filling factors are related to the applied voltages by:
\begin{align*}
& \nu_\text{t}^0=V_\text{TG}C_\text{TG}/e n_{\phi}    \\
& \nu_\text{m}^0=V_\text{BL}C_\text{BL}/e n_{\phi}  \\
& \nu_\text{b}^0=(V_\text{BG}C_\text{BG}-V_\text{BL}(C_{BL}+C_{BG}))/e n_{\phi}
\end{align*}
Although the quantum capacitance of each layer produces noticeable corrections to $\nu_\text{t}^0$ , $\nu_\text{m}^0$ and $\nu_\text{b}^0$ because of the close proximity of adjacent layers, the total filling factor of all three layers \nutotal is determined primarily by $V_\text{TG}$ and $V_\text{BG}$. This is evidenced by nearly \vbl-independent \cp\ peaks corresponding to constant \nutotal\ (Fig. S1a). From the incompressible features in \cg\ (Fig. S1b), we obtain the ratio $C_\text{BL}/C_\text{BG}=26.3$, corresponding to $d_\text{bm}= 2.4$ nm. This allows us to sweep $\nu_\text{t}^0$ and $\nu_\text{m}^0$ at fixed $\nu_\text{b}^0=V_\text{BG}^*C_\text{BG}/e n_{\phi}$ by maintaining $V_\text{BG}=27.3V_\text{BL}+V_\text{BG}^*$. The similar charge gaps between \xtm and \xbm condensates suggest $d_\text{tm}\approx d_\text{bm}$.

Away from integer fillings, the compressibility of a compressible composite Fermi liquid is approximately constant\cite{Yang_prl2021_chemicalpotential}, so the graphene quantum capacitance introduces only a first-order correction to the carrier density. A linear transformation can therefore map the filling factors estimated from the geometric capacitances ($\nu_\text{t}^0$ , $\nu_\text{m}^0$ , $\nu_\text{b}^0$) onto the calibrated filling factors (\nut, \num, \nub). This approach has been applied to the quantum Hall bilayer graphene and shown to give consistent results\cite{Li2019,liu_interlayer_2019,zhang_excitons_2025,zeng_2026_excitonsolid}. Figure S2 shows representative \cp\ and \cg\ data as functions of ($\nu_\text{t}^0$ , $\nu_\text{m}^0$ , $\nu_\text{b}^0$) before transformation and of (\nut, \num, \nub) after the transformation. 

\begin{figure}
\includegraphics[width=0.7\linewidth]{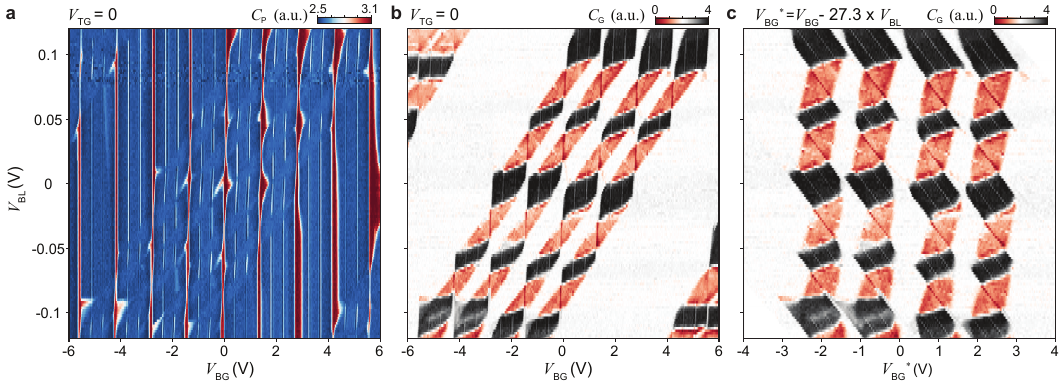}
\caption{{\bf{Capacitance Measurement at \vtg = 0~V.}} \textbf{a,b}, \cp\ (\textbf{a}) and \cg\ (\textbf{b}) as functions of \vbg{} and \vbl{} at \vtg  = 0 V. The four bands in \textbf{b} correspond to the four Landau levels in the bottom layer, while the top layer remains in an integer quantum Hall gap. Their slopes yield the geometric-capacitance ratio $C_\text{BL}/C_\text{BG} = 26.3$. \textbf{c}, Data from \textbf{b} replotted as functions of $V_\text{BG}^*$ and \vbl{} showing the expected bottom-layer response after removal of the gating contribution from \vbl.}
\end{figure}

\begin{figure}
\includegraphics[width=0.7\linewidth]{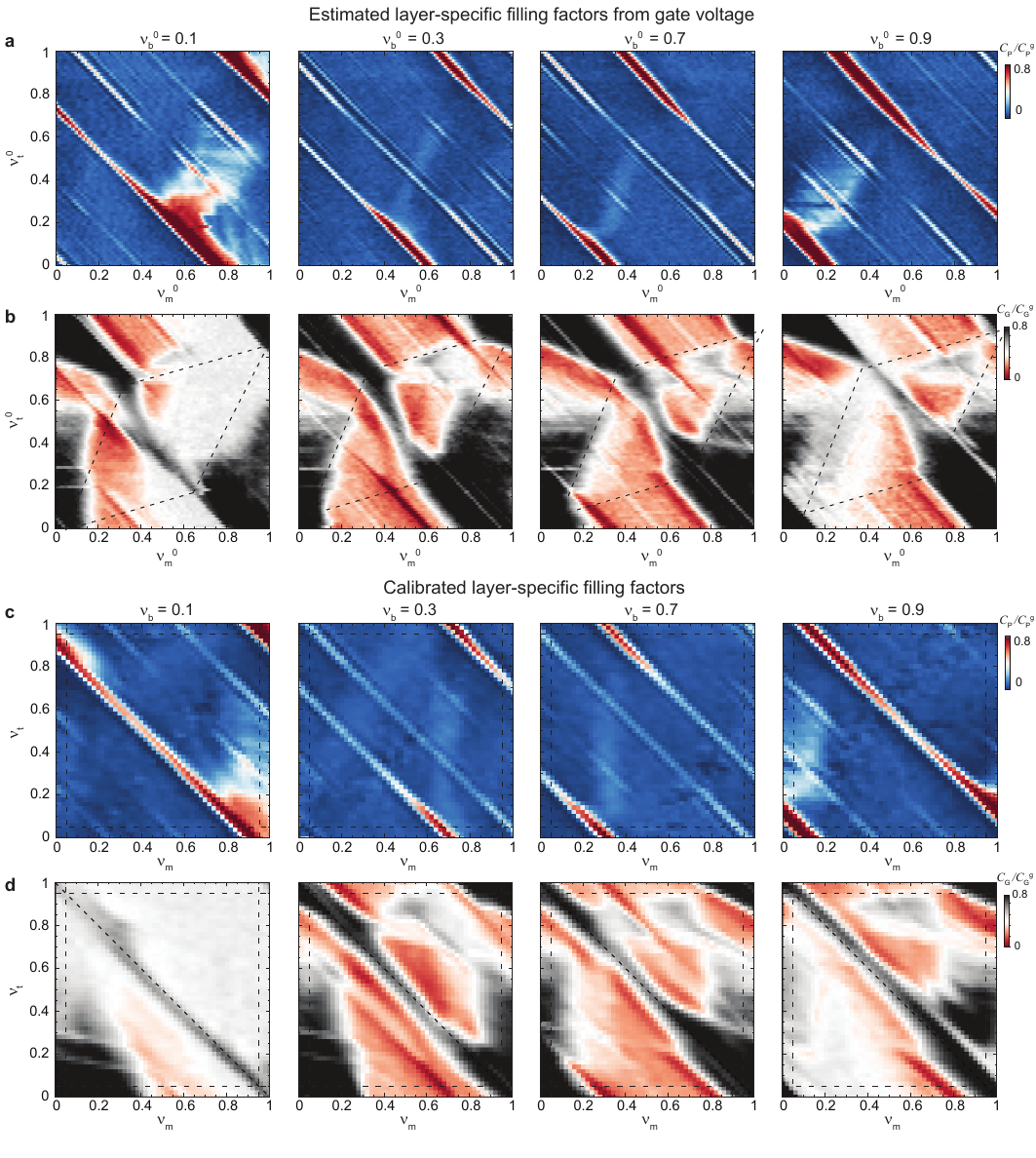}
\caption{{\bf{Transformation of layer-specific filling factors.}} \textbf{a,b}, \cp\ (\textbf{a}) and \cg\ (\textbf{b}) as functions of $\nu_\text{t}^0$ and $\nu_\text{m}^0$ estimated from gate voltages and geometric capacitances, for $\nu_\text{b}^0$= 0.1, 0.3, 0.7 and 0.9. For each $\nu_\text{b}^0$, the region $0<\nu_\text{t}^0, \nu_\text{m}^0<1$ forms a diamond shape because of quantum-capacitance effect (black dashed lines), similar to that in Fig 1g. The diamond also shifts with $\nu_\text{b}^0$ owing to the quantum capacitance. \textbf{c,d}, \cp\ (\textbf{c}) and \cg\ (\textbf{d}) as functions of the calibrated layer-specific filling factors after the linear transformation. The transformation maps each diamond onto a square in the \nut-\num\ plane and aligns the top–middle incompressible feature with \nut+\num=1 for all \nub.}
\end{figure}

We determine the transformation matrix using four consistency checks of \cp\ and \cg\ as functions of the calibrated filling factors \nut, \num, and \nub: (1) the top--middle incompressible state observed in \cg\ aligns with \nut$+$\num$=1$, and the associated \cg\ feature reaches its minimum near the density-balanced point \nut$=$~\num$=1/2$ (Fig.~2b,e and Fig.~S2d); (2) the integer-total-filling incompressible states observed in \cp\ align with $\nu_{\mathrm{total}}=1$ and $2$ throughout the filling-factor space (Fig.~2a,d, Fig.~S2c, and Fig.~S3); (3) the charge gaps of the single-component condensates \xtm, \xbm, and \xtb\ reach minima near their respective density-balanced conditions (Fig.~3a--c); and (4) the filling-factor maps near $\nu_{\mathrm{total}}=1$ and $2$ are approximately related by the layer-resolved particle--hole transformation (\nut,\num,\nub)$\leftrightarrow(1-\nu_\text{t},1-\nu_\text{m},1-\nu_\text{b})$ (Fig.~2g,h and Extended Data Fig.~4).

The fractional-total-filling features were not used to determine the transformation matrix and therefore provide independent consistency checks. These include the incompressible features at the expected fractional total fillings in Fig.~2a,c,d,f and the fractional-total-filling phase diagram in Fig.~2i, including the features associated with $\nut=1/3, \nub=1/3$, and the selected commensurate filling configurations. Their agreement with the expected fractional fillings supports the calibration independently of the observables used in its initial determination.

Away from the layer-polarized integer quantum Hall limits, the extracted integer-$\nu_{\mathrm{total}}$ peak positions deviate from their expected values by $\lesssim0.02$ (Fig.~S3). Accounting conservatively for the finite widths of the layer-resolved calibration features, we estimate an uncertainty of approximately $0.04$ in each calibrated layer filling factor. This uncertainty may slightly shift the precise filling-factor coordinates of the observed features but does not affect their identification or the  conclusions of this work.

\begin{figure}
\includegraphics[width=0.7\linewidth]{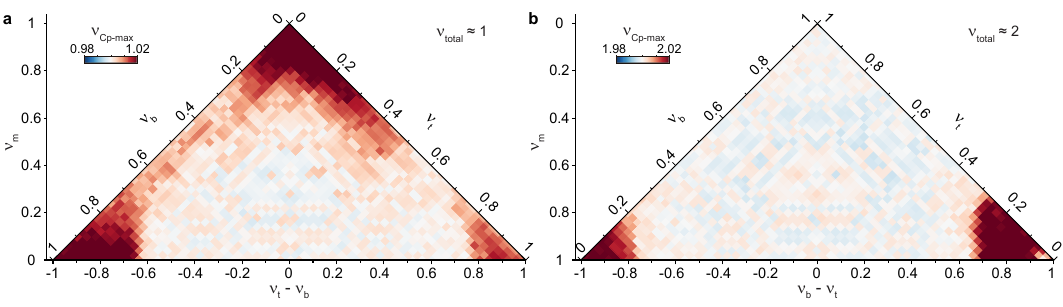}
\caption{{\bf{Filling factor for maximum \cp\ near \nutotal\ = 1 and 2.}} \textbf{a,b}, Peak position $\nu_\text{Cp-max}$ as defined in Extended Data Fig. 1 near \nutotal\ = 1 and \nutotal\ = 2. The results demonstrate consistent filling factor calibration throughout most of the (\nut, \num, \nub) phase space. Deviations from $\nutotal=1$ or 2 occur near the layer-decoupled integer quantum Hall limits because of the strong non-linear charging effect near a robust insulator and the broad $\cp/C_\text{P}^\text{g}$ plateau.}
\end{figure}
\FloatBarrier

\section{Hartree-Fock calculations for the  \nutotal = 1 exciton condensate in a Coulomb-coupled trilayer}

Here we introduce the mean-field theory of Coulomb-coupled trilayer zeroth Landau levels at \nutotal = 1 without spin/valley degrees of freedom. 
Using Hartree-fock approximation, we find spatially uniform ground states with all three layers are partially filled in which the $U(1)\times U(1)$ symmetry is spontaneously.
The voltages in experiments are mapped onto electrochemical potentials ($\mu_t$, $\mu_m$, $\mu_b$) in the Hamiltonian, which control the ground state layer filling factor ($\nu_t$, $\nu_m$, $\nu_b$).
This framework was applied in \cite{83f9-pln9} in the bilayer case and is extended to three layers here.
By enforcing finite momenta of interlayer coherences, we find the stiffness matrix.
We also use the time-dependent Hartree-Fock (TDHF) method to calculate the charge-neutral excitation modes and show the two Goldstone modes with linear dispersion in the long-wavelength limit.

The Hamiltonian 
\begin{equation}
    H = \sum_{l, X} -\mu_l\,c^\dagger_{lX} c_{lX} + \frac{1}{2A}\sum_\textbf{q} \sum_{l\,l^\prime}\sum_{X_1X_2X_3X_4} V_{ll^\prime}(\textbf{q}) \left< X_1 \middle| e^{i\textbf{q}\cdot\textbf{r}} \middle| X_4 \right> \left< X_2 \middle| e^{-i\textbf{q}\cdot\textbf{r}} \middle| X_3 \right> c^\dagger_{l^\prime X_1} c^\dagger_{l X_2} c_{l X_3} c_{l^\prime X_4},
\end{equation}
where $A$ is the sample area, layer index $l\in\{t,m,b\}$, $\mu_l$ is electrochemical potential of layer $l$, guiding center $X$ labels the orbital in the zeroth Landau levels, and $V_{ll^\prime}(\textbf{q})$ is the Fourier transform of dual-gated Coulomb interaction between layer $l$ and $l^\prime$.
\begin{equation}
    V_{ll^\prime}(\textbf{q}) = \frac{2\pi e^2}{\epsilon q} \frac{e^{-qd_{ll^\prime}} \left(1-e^{-2qD_{TG,\text{top}(l,l^\prime)}}\right)\left(1-e^{-2qD_{BG,\text{bot}(l,l^\prime)}}\right)}{1-e^{-2qD_{TG,BG}}},
\end{equation}
where $q$ is the length of $\textbf{q}$, $d_{l,l^\prime}$ is the distance between layer $l$ and $l^\prime$, $D_{TG,BG}$ is the gate-to-gate distance, and the two gate-to-layer distances $D_{TG,\text{top}(l,l^\prime)}$ and $D_{BG,\text{bot}(l,l^\prime)}$ are distances from the upper of the two layers ($l$ an $l^\prime$) to the top gate and from the lower one to the bottom gate, respectively.
For the results shown in the main text, we use gate distances $D_{TG,t} = D_{BG,b} = 30nm$, interlayer distances $d_{tm} = d_{mb} = 3nm$, magnetic field $B = 15T$, and relative dielectric constant $\epsilon = 15$.

For negatively charged electrons and the magnetic field in the $+z$ direction, we define the momentum-resolved density operator
\begin{equation}
    \rho_{ll^\prime}(\textbf{q}) = \frac{2\pi \ell_B^2}{A} \sum_{X} e^{-iq_x(X+\frac{1}{2}q_y\ell_B^2)} c^\dagger_{l^\prime,X+q_y\ell_B^2} c_{l,X}.
\end{equation}
Its expectation value is know as the momentum-resolved density matrix. $\ell_B$ is the magnetic length.

The Hartree-Fock approximation is employed to find ground state density matrices $\left< \rho_{ll^\prime} (\textbf{q}) \right>$ with zero momentum.
To further evaluate the stiffness, we also assign momenta $\textbf{q}_{tm}$ and $\textbf{q}_{mb}$ for the interlayer coherence between the top and middle layer and between the middle and bottom layer;
the top-bottom interlayer coherence momentum is then $\textbf{q}_{tb} = \textbf{q}_{tm} + \textbf{q}_{mb}$.
The intralayer densities remain at zero momentum, and $\textbf{q}_{l^\prime l} = -\textbf{q}_{ll^\prime}$.
\begin{equation} \label{eq:density matrix}
    \rho_{ll^\prime}(\textbf{q}_{tm},\textbf{q}_{mb}) = 
    \begin{pmatrix}
        \rho_{tt}(0) & \rho_{tm}(\textbf{q}_{tm}) & \rho_{tb}(\textbf{q}_{tb})\\
        \rho_{mt}(-\textbf{q}_{tm}) & \rho_{mm}(0) & \rho_{mb}(\textbf{q}_{mb})\\
        \rho_{bt}(-\textbf{q}_{tb}) & \rho_{bm}(-\textbf{q}_{mb}) & \rho_{bb}(0)\\
    \end{pmatrix}.
\end{equation}
At \nutotal=1, the density matrix is constructed from the lowest-energy eigenstate of the Hartree-Fock Hamiltonian
\begin{equation}
\begin{aligned}
    H^{\text{HF}}(\textbf{q}_{tm},\textbf{q}_{mb})_{ll^\prime} &= -\mu_l\delta_{ll^\prime} + H^\text{int}(\textbf{q}_{tm},\textbf{q}_{mb})_{ll^\prime}\\
    &= -\mu_l\delta_{ll^\prime} + \sum_{l_1} D_{ll_1}(0) \left< \rho_{l_1 l_1}(0)\right> \delta_{ll^\prime} - X_{ll^\prime}(q_{ll^\prime}) \left<\rho_{ll^\prime}(\textbf{q}_{ll^\prime})\right> ,
\end{aligned}
\end{equation}
and is obtained through self-consistent iteration.
The coefficients in the Hartree (direct) potential
\begin{equation}
    D_{ll^\prime}(q) = V_{ll^\prime}(\textbf{q}) e^{-\frac{1}{2}q^2\ell_B^2},
\end{equation}
and in the Fock (exchange) potential
\begin{equation}
    X_{ll^\prime}(q) = \int_0^\infty d{k}\, V_{ll^\prime}(k) e^{-\frac{1}{2}k^2\ell_B^2} J_0(kq\ell_B^2).
\end{equation}

The energy of the converged ground state $\rho_{ll^\prime}(\textbf{q}_{tm},\textbf{q}_{mb})$
\begin{equation}
    E(\textbf{q}_{tm},\textbf{q}_{mb}) = -\frac{1}{2} \sum_l \mu_l \left< \rho_{ll}(0) \right> + \frac{1}{2} \text{Tr} \left( H^\text{HF} \ \left<\rho\right> \right),
\end{equation}
is a quadratic form on the parameter space of the two momenta in the long-wavelength limit.
\begin{equation}
    E(\textbf{q}_{tm},\textbf{q}_{mb}) = E(0,0) + \frac{1}{2} \begin{bmatrix} q_{tm} & q_{mb} \end{bmatrix} \begin{bmatrix}
        \rho_{s\, 11} & \rho_{s\, 12}\\
        \rho_{s\, 21} & \rho_{s\, 22}\\
    \end{bmatrix}
    \begin{bmatrix} q_{tm} \\ q_{mb} \end{bmatrix}.
\end{equation}
The stiffness matrix $\rho_s$, is real-symmetric and positively semi-definite; therefore, its eigenvalues are positive, and its eigenvectors can be chosen to have real components.
Each eigenvector represents the a mode of "trilayer counterflow" involving all three layers with zero net current, and can be viewed as the mixture of the top-middle counterflow and middle-bottom counterflow with a certain ratio. 
For the real eigenvector $\left[ a_{tm}\ a_{mb} \right]^\text{T}$, this mixing ratio of the two counterflows is represented with the tangent of angle $\theta_stiff$ modulo $\pi$.
\begin{equation}
    \tan \theta_{stiff} = a_{tm} / a_{mb}.
\end{equation}

In TDHF calculation, we solve the excitation modes of the self-consistent ground state with $\textbf{q}_{tm}=\textbf{q}_{mb}=0$ by finding eigenmodes of the susceptibility kernel $L_{\alpha,\beta}(\textbf{k})$
for the density-density linear response within the mean-field approximation regime for Coulomb interaction.
The Greek subscripts denotes pairs of layer indices $(l,l^\prime)$ that label the full nine components of the density operator in Eq.~\ref{eq:density matrix}, and $\textbf{k}$ stands for the momentum of excitations.
\begin{equation}
\begin{aligned}
    L_{\alpha\beta}(\textbf{k}) & = \delta_{\alpha \beta} (\mu_{l^\prime_{\alpha}} - \mu_{l_{\alpha}})
    - \delta_{l_\alpha l_\beta} H^\text{int}(0,0)_{l^\prime_\alpha l^\prime_\beta}
    + \delta_{l^\prime_\alpha l^\prime_\beta} H^\text{int}(0,0)_{l_\alpha l_\beta}\\
    &\ \ + \sum_\gamma \left( \delta_{l_\alpha l_\gamma} \big< \rho_{l^\prime_\gamma l^\prime_\alpha} \big> 
    - \delta_{l^\prime_\alpha l^\prime_\gamma} \left< \rho_{l_\alpha l_\gamma} \right> \right)
    \left( \delta_{l_\beta l^\prime_\beta} \delta_{l_\gamma l^\prime_\gamma} D_{l_\gamma l_\beta}(k) 
    - \delta_{\beta\gamma} X_{l_\beta l^\prime_\beta}(k) \right).
\end{aligned}
\end{equation}

This TDHF matrix is expressed in the full density basis instead of being projected onto the "particle-hole" excitation and "hole-particle" de-excitation sectors.
Therefore, five of its nine eigenmodes are unphysical, one in the "hole-hole" sector and four in the "particle-particle" sector.
These unphysical modes exhibit $\textbf{k}$-independent eigenvalues that equals to the differences of the eigenenergies of the mean-field Hamiltonian.
The remaining four physical eigenvalues occur in two pairs $\pm \omega_1(\textbf{k})$ and $\pm \omega_2(\textbf{k})$, abd reflect the excitation energies of the two collective Goldstone modes.
At the long-wavelength limit, they exhibit linear dispersion where the slopes define the first sound speeds $v$ of the exciton superfluids.
When setting both order parameter phases to zero in the ground state density matrix, the corresponding eigenvectors have real layer-diagonal components ($a_{tt}$, $a_{mm}$, $a_{bb}$) that sum up to zero and describe the ratio of charge transfer among the layers.
We similarly decompose the charge transfer into the mixture of transfers between top-middle layers and middle-bottom layers, and their ratio is represented by
\begin{equation}
    \tan \theta_{tdhf} = -a_{tt}/a_{bb}.
\end{equation}

The stiffness and TDHF calculations give two eigenmodes of condensation in combinations of the top-middle layer exciton $X_{tm}$ and middle-bottom layer exciton $X_{bm}$, 
providing complimentary decompositions of the coupled two-component condensates.
The stiffness eigenmodes suggest collective excitons in condensation that only have spatial phases winding.
By contrast, the TDHF superfluid phonon excitations the eigenstates of time-dependent Hamiltonian and contain spatial and temporal oscillations of both condensate phases and layer densities.
The sound velocities depend jointly the stiffness and the layer density susceptibility. 
Because the layer susceptibility is not identical matrix in the two components, the $\theta$ from the two calculations need not coincide.

\bibliography{YZ_ref}%
